\documentstyle[12pt]{article}

\def\RR{\hbox{I\kern -2pt R} }
\def\NN{\hbox{I\kern -2pt N} }
\def\U1{\hbox{1\kern -3pt l} }

\def\EE{\hbox{I\kern -2pt E} }

\def\BB{\hbox{$\cal B$} }
\def\Rt{\hbox{$\cal R$} }

\def\Pn{\hbox{${\cal P}_N$} }

\def\Lm{\displaystyle\lim_{N \rightarrow \infty} {1\over N}}

\font\twelve=cmbx10 at 15pt
\font\ten=cmbx10 at 12pt

\begin{document}

\begin{titlepage}

\begin{center}

{\ten Centre de Physique Th\'eorique - CNRS - Luminy, Case 907}

{\ten F-13288 Marseille Cedex 9 - France }

{\ten Unit\'e Propre de Recherche 7061}

\vspace{1 cm}

{\twelve EXACT LYAPUNOV EXPONENT}

{\twelve FOR INFINITE PRODUCTS}

{\twelve OF RANDOM MATRICES}

\vspace{0.3 cm}

{\bf R. LIMA and M. RAHIBE}\footnote{The Benjamin Levich Institute,
The City College of The City University of New-York Steinman Hall T-1M
New York, N.Y. 10031}

\vspace{2,2 cm}

{\bf Abstract}

\end{center}

In this work, we give a rigorous explicit formula for the  Lyapunov
exponent for some binary infinite products of random $2\times 2$
real matrices. All these products are constructed using only two
types of matrices, $A$ and $B$, which are chosen according to a
stochastic process. The matrix $A$ is singular, namely its
determinant is zero. This formula is derived by using a particular
decomposition  for the matrix $B$, which allows us to write the
Lyapunov exponent as a sum of convergent  series. Finally, we show
with an example that the Lyapunov exponent is a discontinuous
function of the given parameter.

\vspace{2,8 cm}

\noindent December 1993

\noindent CPT-93/P.2974

\bigskip

\noindent anonymous ftp or gopher: cpt.univ-mrs.fr

\end{titlepage}

{\large\bf 1) Introduction }

\vspace{0.5cm}

The product of random matrices  appears in the study of
disordered systems [1] as well as in the context of dynamical systems
[2].  The Lyapunov exponents are one of the tools to study these
products. They are related to physical quantities in disordered
systems [3]. For example, in the thight binding model or
Anderson model [3], the localisation length of the wave function
is proportional to the inverse of the Lyapunov exponent. In
dynamical systems theory, products of random matrices often arise
as a non trivial approximation wich mimics strongly chaotic behavior
in deterministic systems.

\vspace{0.5cm}

In spite of the important and numerous  results obtained in the
theory of random matrices [4,5,6], there is no general method for
calculating the Lyapunov exponents [7,8,9], although for a
few examples there is an explicit formula [10,11].

\vspace{0.5cm}

In this paper, we present some examples of products of random
matrices, where we were able to determine the Lyapunov exponents  as a
sum  of explicitly convergent series. In all these examples, we deal
with infinite binary random products, built with only two types of
$2\times2$ real matrices, $A$ and $B$ which are chosen
according to  a stochastic Bernoulli or Markovian process  and one of
the matrices, say $A$,  is singular,i.e. its determinant is zero.
Markovian processes are chosen to mimic the correlations existing in
dynamical systems exhibiting weak chaos [12]. The key of our study is
the use of a particular  decomposition when the matrix $B$ is
non-singular (but if B is singular, the calculation can be done
directly). The Bernoulli case was studied by Pincus [13], who gave
assymptotic results for the Lyapunov exponent, but which do not enable
an actual computation, see remark 2.3 below (also Derrida and Hilhorst
[14] have obtained a similar formula for a particular product).
Instead, we give an explicit form of the terms of that series.  In
some cases, this permits either an explicit summation of the series,
or   a control of the convergence,  therefore leading to approximative
results. Furthermore, we could show that the convergence of the series
is exponentially fast.

\newpage

The Markovian products are treated in the same way as the Bernoulli
one.

\vspace{0.5cm}

The paper is organized as follows. In section 2, we derive the general
formula for the largest Lyapunov exponent $\gamma$. Section 3 is
devoted to the simple case where $B$ is a singular matrix. In section
4 we perform the decomposition of $B$ wich, in turn, leads to the
decompostion of $B^n$ and
 in section 5 we derive the expression of $\gamma$. In
the last section, we analyze the continuity of $\gamma$ as
function of parameters.

\vspace{0.5cm}

{\large\bf 2) Lyapunov exponent, general formula }

\vspace{0.5cm}

We consider  $A$ and $B$, $2 \times 2$ real matrices, $A$ and $B$,
where $A$ is a singular matrix. Now let $\Pn$ :
$$\Pn = X_N X_{N-1}\cdots \cdots X_2
X_1$$
be a binary product of $N$ matrices, where the matrix $X_i$ is either
$A$ or  $B$, the choice being made by a stochastic process. In the
Bernoulli case, we have $X_i = A$ with probability $p$ $( 0 < p < 1)$
and $X_i = B$ with probability $q = p - 1 $,  $(i \geq 1)$. In the
Markovian case,  the transition probabilities are given by  $$ \cases
{ Pr( X_{n+1} = A / X_n = B ) = p_1 \cr Pr( X_{n+1} = B / X_n = A ) =
p_2 \cr Pr( X_{n+1} = B / X_n = B ) = 1-p_1 = q_1 \cr Pr( X_{n+1} = A
/ X_n = A ) = 1-p_2 = q_2\cr }\eqno(2.1)$$ and $$\cases { Pr( X_1 = A
) =
\displaystyle{\frac {p_1}{p_1 + p_2}  = p_0} \cr Pr( X_1 = B ) =
\displaystyle{\frac {p_2}{p_1 + p_2}  = q_0} \cr    }\eqno(2.2)$$
with $0 < p_1 < 1$ and $0 < p_2 <1$.

By definition, the Lyapunov exponent $\gamma$ is  $$\gamma =
\Lm \log {\parallel \Pn \parallel}.\eqno(2.3)$$

$\gamma$ is independent of the choice of the norm $\parallel
\parallel$.

\vspace{0.5cm}

Since  the matrix $A$ is singular,  it can be written by a
change of  basis in one of the following form :

$$ A = \pmatrix { \lambda & 0 \cr 0    & 0 \cr} \eqno(2.4)$$  or,
$$A = \pmatrix { 0       & \lambda  \cr 0      & 0 \cr}
\eqno(2.5)$$

If $A$ is of the type $(2.5)$, the Lyapunov exponent is $\gamma =
-\infty$. This is straightforward to show, since  $ A^2
= 0 $.

\vspace{0.5cm}

Therefore we suppose,  without loss of generality, that $A$ has the
form $(2.4)$, i.e. $A = \pmatrix { \lambda & 0 \cr 0    & 0 \cr}$.

\vspace{0.5cm}

If we write $B^n$ in the form $B^n =\pmatrix { b_{11}(n)      &
b_{12}(n) \cr
               &            \cr b_{21}(n)      & b_{22}(n) \cr} $,
then the result obtained by Pincus $[13]$, in the
Bernoulli case, follows :
$$\gamma  = p \log\mid{\lambda}\mid +
 \sum_{n=1}^{\infty}{ p^2 (1-p)^{n} \log\mid b_{11}(n) \mid},
\eqno(2.6)$$

Where $p^2(1-p)^n$ is the probability to obtain the
subproduct $A B^n A$. Even if $(2.6)$ was proved only in the Bernoulli
case [13], the same argument extends to the Markovian case, leading
to the following proposition.

\vspace{0.5cm}

{\bf Proposition 2.1:  } Let \{ $\Pn$ \} be an infinite product of
random matrices satisfying the Markovian distribution law $(2.1)$ and
$(2.2)$, where
$A$ is a singular matrix given by $(2.4)$ and $B$ is general.

Then

$$\gamma =   {p_1 \over {p_1
+ p_2}} \log\mid{\lambda}\mid + \sum_{n=1}^{\infty} { p_0 p_1
p_2{q_1}^{n-1} \log{\mid b_{11}(n) \mid} } \eqno(2.7)$$
where $b_{11}(n)$ will be explicitly computed in section 4.

\vspace{0.5cm}

{\bf Remark 2.2:  } The proof of this proposition  is
analogous to that given in [13] once we notice that the probability to
find the product $A B^n A$ is $p_0 p_1 q_1^{n-1}$ in the Markovian
case whereas it is $p^2 ( 1 - p )$ in the Bernoulli case.

\vspace{0.5cm}

{\bf Remark 2.3:  } Notice that, in order to give an explicit value
for $\gamma$, from $(2.6)$ for the Bernouli case or $(2.7)$ for the
markovian case, the key point is the computation of  $b_{11}(n)$, a
problem which was not addressed in [13]. This is the object of the
next sections

According to $(2.7)$, it is possible to study the Bernoulli and the
Markovian cases in the same manner, by writing
$$\gamma =
Pr(A) \log\mid{\lambda}\mid +  L \sum_{n=1}^{\infty}{x^{n-1}
\log\mid b_{11}(n) \mid}  \eqno(2.8)$$
where
$$\cases {
\displaystyle{ Pr(A) = p} \cr \displaystyle{ }\cr
\displaystyle{ L =
p^2 (1-p)  }\cr \displaystyle{ }\cr \displaystyle{ x = 1-p } \cr
}\eqno(2.9)$$
in the Bernoulli case, and
$$\cases { \displaystyle{ Pr(A) = {p_1 \over {p_1 + p_2}} }\cr
\displaystyle{ L = {{p_1^2 p_2} \over {p_1 + p_2}}  }\cr
\displaystyle{ }\cr \displaystyle{ x = q_1} \cr   }\eqno(2.10)$$
in the Markovian case.

\vspace{0.5cm}

{\large\bf 3) The case of two singular matrices }

\vspace{0.5cm}

In the case where $B$ is singular, it is easy to
calculate $b_{11}(n)$. Indeed, as explained, above we
have two different cases,

either
$$B= Q^{-1} \pmatrix { \lambda_b & 0 \cr 0    & 0
\cr} Q$$  or, $$B= Q^{-1} \pmatrix { 0    & \lambda_b\cr 0    & 0
\cr} Q$$ where $Q$ is an invertible matrix and $Q^{-1}$ is its
inverse.

In the former case $\gamma = - \infty$, by the same
argument as when A is given by $(2.5)$.

\vspace{0.5cm}

In the latter case, if $Q$ is written as : $Q = \pmatrix{
q_{11} & q_{12}\cr q_{21}  & q_{22} \cr }$, then we have,  for $n \geq
1$, $b_{11}(n) = {{\lambda}_b}^n q_{11} q_{22}$. And it is then
easy to obtain $\gamma$; for a Bernoulli product, we have
$$ \gamma = {\gamma}_B = p \log\mid{\lambda}\mid + (1-p)
\log\mid{\lambda}_b\mid  + p (1-p) \log{\mid{b_{11} \over
Tr(B)}\mid} \eqno(3.1)$$
and for a Markovian product, we have
$$  \gamma =
{\gamma}_M = {p_1 \over {p_1 + p_2}}  \log\mid{\lambda}\mid  + {p_2
\over {p_1 + p_2}}  \log\mid{\lambda}_b\mid  + {{p_1 p_2} \over {p_1
+ p_2} } \log\mid{b_{11} \over Tr(B)}\mid \eqno(3.2)$$
where
$b_{11}$ denotes the first entry of the matrix $B$, and $Tr(B)$ is the
trace of $B$.

\vspace{0.5cm}

In the formulae $(2.11)$ and $(2.12)$, we notice a non-linear term,
originating in the non-commutativity of the matrices $A$ and $B$.

\vspace{0.5cm}

{\large\bf 4) Normal form and computation of $B^n$ }

\vspace{0.5cm}

When $B$ is non-singular, we introduce a decomposition,
which enables  us to determine $b_{11}(n)$ for all positive integers
$n$.

\vspace{0.5cm}

Let $B$ be a real, non-singular $2 \times 2$ matrix, which we write
as
$$B
= \mid \det B \mid \Rt (-\varphi ) \BB \Rt (\varphi )\eqno(4.1)$$
where $\det B$ is the determinant of $B$, $\Rt ( \varphi )=\pmatrix{
\cos\varphi & -\sin\varphi \cr \sin\varphi & \cos\varphi \cr}$ is a
matrix rotation in the plane with an angle $\varphi$ and $\Rt
(-\varphi )$ is its inverse.  The angle $\varphi$ is determined
by
$$ \tan( 2 \varphi ) = {b_{22} - b_{11} \over b_{12} + b_{21}}
\eqno(4.2)$$
where $b_{ij}$ are the entries of $B$.

\vspace{0.5cm}

We call the matrix $\BB$, the normal form of  $B$. We have four
different types of normal forms $\BB$, depending  on the sign of the
determinant and on the eigenvalues of $B$, which can be real,
non-degenerated eigenvalues, real degenerated eigenvalues and
conjugate complex eigenvalues.

\vspace{0.5cm}

We now define the quantity : $$\rho = \sqrt{\mid{{l_2} \over {l_1}}
\mid}\eqno(4.3)$$  where  $$ l_1 = b_{12} \sin^2(\varphi) - b_{21}
\cos^2(\varphi) +
 {1 \over 2} (b_{11} - b_{22}) \sin(2 \varphi)$$ and $$ l_2 = b_{12}
\cos^2(\varphi) - b_{21} \sin^2(\varphi) -  {1 \over 2} (b_{11} -
b_{22}) \sin(2 \varphi)$$

$\rho$ will be used in the expression of $\BB$.

\vspace{0.5cm}

{\bf Remark 4.1:  } Replacing $B$ by $C = {\displaystyle{1 \over {\det
B}}B}$, accroding to $(2.3)$, $\gamma$ is shifted by an additive
constant, $
\log \mid \det B \mid$. Therefore, we can assume without loss of
generality, that $\mid \det B \mid = 1$.

\vspace{0.5cm}

We now consider four different cases.

{\bf Case I : $B$ is hyperbolic symplectic }

\vspace{0.5cm}

The eigenvalues of $B$ are real non-degenerated and have the same
sign,
$\lambda _{1,2} = \epsilon \exp{\pm \sigma}$ with $\sigma$ non zero
and  $\epsilon = 1$ or $ -1$ ; then, we get
$$\BB = \epsilon
\pmatrix{  \cosh {\sigma}                 & {\rho} \sinh {\sigma} \cr
\displaystyle{1 \over {\rho}}\sinh {\sigma}   &  \cosh {\sigma}
\cr}\eqno(4.4)$$

\vspace{0.5cm}

{\bf Case II : $B$ is hyperbolic, non-symplectic }

\vspace{0.5cm}

In this case the eigenvalues of $B$ are also real, non-degenerated but
they have opposite signs, namely  ${\lambda}_{1} = \exp{\sigma}$ and
${\lambda}_{2} = -\exp{-\sigma}$ or $-{\lambda}_1$ and $-{\lambda}_2$.
Although we have two possible forms for $\BB$,
we can deduce one from the other  by changing the sign of $\varphi$.
We therefore retain only one form for $\BB$:
$$\BB = \pmatrix{  \sinh
{\sigma}                  & {\rho} \cosh {\sigma} \cr \displaystyle{1
\over {\rho}}\cosh {\sigma}   &  \sinh {\sigma}  \cr}\eqno(4.5)$$

\vspace{0.5cm}

{\bf Case III : $B$ is parabolic }

\vspace{0.5cm}

The eigenvalues of $B$ are given by
$\lambda _{1,2} = \epsilon $, ($\epsilon = 1$ or $-1$) and we have
either
$$\BB = \epsilon \pmatrix{ 1   &   \epsilon l_2  \cr 0   &    1
\cr},\eqno(4.6)$$
or
$$\BB = \epsilon \pmatrix{ 1     &    0  \cr -\epsilon
l_1   &    1   \cr}\eqno(4.7)$$

\vspace{0.5cm}

Notice that   we can pass from $(4.7)$ to $(4.6)$ by a rotation in the
plane with an angle $\displaystyle{\pi \over 2}$, which is
equivalent to a change of the sign of $\varphi$. In the following,
we define the normal form of a parabolic matrix as given by $(4.6)$.

\vspace{0.5cm}

{\bf Case IV : $B$ is elliptic }

\vspace{0.5cm}

The eigenvalues of $B$ are complex conjugate,
$\lambda _{1,2} = \exp{\pm i\sigma}$, in which case
 $$\BB = \pmatrix{
\cos {\sigma}                  & -{\rho} \sin {\sigma} \cr
\displaystyle{1 \over {\rho}}\sin {\sigma}   &  \cos {\sigma}
\cr}.\eqno(4.8)$$

\vspace{0.5cm}

Therefore, if $B$ is a non-singular matrix, we can write it in
the form given by $(4.1)$ where $\BB$ is given by  one of the
expressions
$(4.4)$, $(4.5)$, $(4.6)$ or $(4.8)$.

\vspace{0.5cm}

We can now easily compute $B^n$ in each of the previous cases. We
summarize the result in the following proposition.

\vspace{0.5cm}

{\bf Proposition 4.2:  } Let $B$ be a non-singular matrix which a
normal form $\BB$ defined by $(4.1)$

Then

$$B^n
= \Rt (-\varphi ) {\BB}^n \Rt (\varphi ).\eqno(4.9)$$
If $B$ is symplectic hyperbolic, given by $(4.4)$, then
$${\BB}^n = {\epsilon}^n \pmatrix{ \cosh {n\sigma}                &
{\rho}
\sinh {n\sigma} \cr \displaystyle{1 \over {\rho}}\sinh {n\sigma}  &
\cosh {n\sigma}  \cr}.\eqno(4.10)$$

If $B$ is non-symplectic hyperbolic, given by $(4.5)$, then  either
$${\BB}^n = \pmatrix{ \cosh {n\sigma}                  & {\rho}
\sinh {n\sigma} \cr \displaystyle{1 \over {\rho}}\sinh {n\sigma}   &
\cosh {n\sigma}  \cr}\eqno(4.11)$$
if n is odd; or
$${\BB}^n =
\pmatrix{ \sinh {n\sigma}                  & {\rho} \cosh {n\sigma}
\cr \displaystyle{1 \over {\rho}}\cosh {n\sigma}   &  \sinh
{n\sigma}  \cr}\eqno(4.12)$$
if n is even.

If $B$ is parabolic, given by $(4.6)$, then
$${\BB}^n = {\epsilon}^n \pmatrix{
1   &   n  {\epsilon}^n l_2  \cr 0   &    1   \cr}.\eqno(4.13)$$
If $B$ is elliptic, given by $(4.8)$, then
$${\BB}^n = \pmatrix{ \cos
{n\sigma}                  & -{\rho} \sin {n\sigma} \cr
\displaystyle{1
\over {\rho}}\sin {n\sigma}   &  \cos {n\sigma}  \cr}\eqno(4.14)$$

\vspace{0.5cm}

{\large\bf 5) Lyapunov Exponent : explicit formulae}

\vspace{0.5cm}

We are now ready to perform  the analysis of the formula
$(2.8)$ in the case where the matrix $B$ is non-singular.

\vspace{0.5cm}

As previously, without loss of generality,  we suppose that $\mid
\det B\mid = 1$.

\vspace{0.5cm}

We now give the value of the largest Liapunov exponent in each of the
four cases treated in the previous section.

\vspace{0.5cm}

{\bf Case I : B is an hyperbolic symplectic matrix }

For $n \geq 1$, we have  $$b_{11}(n) = {\epsilon }^n [
\cosh({n\sigma}) +
 \cosh(\alpha) \sinh({n\sigma})  \sin(2 \varphi)  ]\eqno(5.1)$$
where
$$\cosh \alpha = {1 \over 2} ( \rho + {1 \over \rho} )
\eqno(5.2)$$

In $(5.1)$, we can  take $\sigma > 0$. Indeed, if $\sigma$ is
negative, we change the sign of $\varphi$ and use $-\sigma$ instead
of $\sigma$.

\vspace{0.5cm}

Notice  first that if it exists an integer $n_0
\geq 1$ such that  $b_{11}(n_0) = 0$ then $\gamma = - \infty$.

\vspace{0.5cm}

The condition $b_{11}(n_0) = 0 $ is equivalent to $$ \cosh \alpha = -
{1
\over{ \sin(2 \varphi) \tanh(n_0\sigma )}}$$
and therefore we can
construct some products for wich $\gamma = - \infty$.

\vspace{0.5cm}

On the contrary, if we suppose that for all integers $n \geq 1$,
$b_{11}(n)
\neq 0$, and we define the quantities $$\delta = 1 +
\sin(2\varphi)\cosh(\alpha)\eqno(5.3)$$
and
$$\tau = { {1 - \sin(
2\varphi )\cosh( \alpha )} \over {1 + \sin( 2\varphi )\cosh( \alpha
)} } \eqno(5.4)$$

We will use $\tau$ and $\delta$ in the expression of $\gamma$.

\vspace{0.5cm}

We now distinguish two cases.

\vspace{0.5cm}

{\hspace{0.5cm}{\bf a)} {\hspace{0.1cm}} $ \delta = 0$ :

A straithforward calculation gives
$$\gamma = {\gamma}_B =  p
\log\mid{\lambda}\mid - (1-p) \sigma \eqno(5.5)$$ for the Lyapunov
exponent in the Bernoulli case, and $$\gamma = {\gamma}_M = {p_1
\over {p_1 + p_2}} \log\mid{\lambda}\mid - {p_2 \over {p_1 + p_2}}
\sigma \eqno(5.6)$$
for the Markovian case.

\vspace{0.5cm}

{\hspace{0.5cm}{\bf b)} {\hspace{0.1cm}} $\delta \neq 0$ :

Depending on whether $\tau$ is zero or not, we have different
expressions for $\gamma$.

\vspace{0.5cm}

{\bf i)} $\tau = 0$ : we easily obtain  the largest Lyapunov exponent
for a Bernoulli product:

$$\gamma = {\gamma}_B =  p \log\mid{\lambda}\mid + (1-p) \sigma
,\eqno(5.7)$$
and for a Markovian product:
$$\gamma = {\gamma}_M =
{p_1 \over {p_1 + p_2}} \log\mid{\lambda}\mid  + {p_2 \over {p_1 +
p_2}} \sigma \eqno(5.8)$$

\vspace{0.5cm}

{\bf ii)} $\tau \neq 0$ : Here we obtain $\gamma$ as a convergent
series given by
$$
\begin{array}{ll} \gamma & =  Pr(A) \log\mid{\lambda}\mid  +  { L
\over {(1-x)^2}} \sigma  \\
 & + { L \over {1-x}} \log{\mid{\delta}\mid \over 2}
 + L \sum_{n=1}^{\infty}{x^{n-1}  \log\mid{1 + \tau
e^{-2n\sigma}}\mid}.
  \end{array}\eqno(5.9)
$$
where $L$ is defined by (2.9) and (2.10).

Notice that if ${\gamma}_N$ is the sum of the N-th first terms in
$(5.9)$ of
$\gamma$, the error is $$\mid{ \gamma - {\gamma}_N }\mid  \leq \mid
\tau\mid  \exp( -2(N+1)\sigma ) L x^{N+1} \eqno(5.10)$$
and therefore it is exponentially small since $0 < x < 1$.

\vspace{0.5cm}

{\bf Case II : B is an hyperbolic no symplectic matrix }

\vspace{0.5cm}

This case is similar to the previous one with only a slight difference
concerning
$(5.9)$. By applying the  decomposition given by $(4.1)$ we obtain
$$b_{11}(n) = {{1 +
\sin(2\varphi)\cosh(\alpha)} \over 2} e^{n\sigma} + (-1)^n {{1 -
\sin(2\varphi) \cosh(\alpha)} \over 2} e^{-n\sigma}.\eqno(5.11)$$

Again, we can suppose that $\sigma > 0$ without loss of generality.
And, as above,  if there exists an $n_0 \geq 1$ such that $b_{11}(2
n_0) = 0$ or
$b_{11}(2 n_0 - 1) = 0$, then $\gamma = -\infty$.

\vspace{0.5cm}

Recall that $\delta$ and $\tau$ are defined by $(5.3)$ and $(5.4)$.

\vspace{0.5cm}

If $\delta = 0$ or if $\delta \neq 0$ but $\tau = 0$, the largest
Lyapunov exponent is given by the same expressions as in the previous
case: $(5.5)$ and $(5.6)$, or $(5.7)$ and $(5.8)$.

\vspace{0.5cm}

Instead, if $\delta \neq 0$ and $\tau \neq 0$, we obtain
$$
\begin{array}{ll} \gamma & =   Pr(A) \log\mid{\lambda}\mid + { L
\over {(1-x)^2}} \sigma + { L \over {1-x}} \log\mid{{\delta}\mid
\over {2}} \\
 & +    L \sum_{n=1}^{\infty}{x^{n-1}  \log\mid{1 + (-1)^n \tau
e^{-2n\sigma}}\mid}.
\end{array} \eqno(5.12)
$$

This serie is convergent, indeed we have
$$\mid{
\gamma  - {\gamma}_N }\mid  \leq   L  x^{N+1} \Delta \eqno(5.13)$$
where $$\Delta = max( \mid\log\mid{1 + (-1)^{N+1}
e^{-2(N+1)\sigma}}\mid\mid , \mid\log\mid{1 + (-1)^{N+2}
e^{-2(N+2)\sigma}}\mid\mid).\eqno(5.14)$$

\vspace{0.5cm}

{\bf Case III : B is a parabolic matrix }

\vspace{0.5cm}

The normal form of $B$ is
$$\BB = \epsilon \pmatrix{ 1 & b \cr 0 &
1}$$
and thus, for $n \geq 1$,
$$b_{11}(n) = {\epsilon }^n [ 1 +  {1
\over 2} n b \sin(2 \varphi)].\eqno(5.15)$$

We suppose that $b_{11} \neq 0$ for all integers $n \geq 1$. The
trivial case where $\sin (2\varphi ) = 0$ corresponds to an
infinite product of diagonal and triangular matrices. In this case,
$$\gamma = Pr(A) \log \mid\lambda\mid .$$

\vspace{0.5cm}

When, instead,  $\sin (2\varphi ) \neq 0$, we have, for a Bernoulli
product:
$$
\begin{array}{ll} \gamma & =  p \log\mid{\lambda}\mid + p
(1-p)  \log \mid{{b \sin(2\varphi)} \over 2 }\mid  \\
 & +  p^2 (1-p)\sum_{n=1}^{\infty}{(1-p)^{n-1}   \log\mid{ n + {1
\over{ n  b \sin(2 \varphi)} }}\mid }   \end{array}\eqno (5.16)$$

and, for a Markovian product,
$$
\begin{array}{ll} \gamma & =  {p_1 \over
{p_1 + p_2}} \log\mid{\lambda}\mid + {{p_1 p_2} \over {p_1 + p_2}}
\log \mid {{b \sin(2\varphi)} \over 2 }\mid  \\
 & +  {{p_1^2 p_2} \over {p_1 + p_2}}
\sum_{n=1}^{\infty}{(1-p_1)^{n-1}  \log\mid{ n + {1 \over{ b \sin(2
\varphi)} }}\mid }.
\end{array}\eqno (5.17)$$

These  series are convergent since we have $$\mid{ \gamma -
{\gamma}_N }\mid  \leq  \mid \log\mid{1 + {2 \over {n b
\sin(\varphi)}}}\mid\mid    \ \ \  \mid {1 \over {1-x}} + {{N+1}
\over (1-x)^2}\mid    x^N
 \eqno(5.18)$$ and $0 < x < 1$.

\vspace{0.5cm}

{\bf Case IV : B is an elliptic matrix }

\vspace{0.5cm}

In this case we also obtain  series for $\gamma$, but it
is very difficult, in general, to evaluate the rest of the
corresponding  partial sums. Indeed, in this situation , as expected,
the summability of the series in the expression of $\gamma$ is related
to the arithmetic properties of $\sigma$, when the latter is
irrational ($mod$ \  $2\pi$). On the another hand, if $\sigma$ is
rational ($mod$ \  $2\pi$), $\gamma$ may be given as a sum of a finite
number of terms.

As above, by using the decomposition $(4.1)$, we obtain
$$b_{11}(n) =
\cos({n\sigma}) +  \sinh (\alpha)  \sin({n\sigma})  \sin(2\varphi)
\eqno(5.19)$$ where
 $$\sinh (\alpha) = {1 \over 2} ( \rho - {1 \over \rho })
\eqno(5.20)$$
with $\alpha  \in {\RR}$.

If $b_{11}(n_0) = 0$ for some positive integer $n_0$, then $\gamma =
- \infty$.

\vspace{0.5cm}

Suppose now that $b_{11}(n) \neq 0$, for all positif integers $n$,
and $\sigma$ is rational ($mod$ \  $2\pi$), i.e. $\sigma =
\displaystyle{r
\over s} 2\pi$ where $r,s \in {\NN}^*$ are irreducible. Thus
$$\gamma = p \log\mid{\lambda}\mid +  {p^2 \over {1 - (1-p)^s}}
 \sum_{j=1}^{s-1}{(1-p)^j  \log\mid{b_{11}(j)}\mid}\eqno(5.21)$$
for
a Bernoulli product,

\noindent and
$$\gamma = {p_1
\over {p_1 + p_2}} \log\mid{\lambda}\mid +  {{p_1^2 p_2} \over {(p_1
+ p_2)(1 - (1-p_1)^s)}}
 \sum_{j=1}^{s-1}{(1-p_1)^{j-1}
\log\mid{b_{11}(j)}\mid}\eqno(5.22)$$
where $$b_{11}(j) =
\cos({{2jr} \over s}\pi) + \sinh\alpha  \sin(2\varphi) \sin({{2jr}
\over s}\pi) \eqno(5.23)$$
for a Markovian product.

\vspace{0.5cm}

{\large\bf 6) One parameter family of products: an example.  }

\vspace{0.5cm}

One may wander if a limit procedure may permit the computation of
$\gamma$  in the elliptic case for irrational $\sigma$ from the
corresponding formulae
$(5.21)$ or
$(5.23)$, when $\sigma$ is rational.

\vspace{0.5cm}

More generally, the continuity of the largest Lyapunov exponent as a
function of the amplitude of the disorder in disordered systems is
known as being an important issue. In this section, we illustrate the
results of the previous sections by giving an example where, indeed,
$\gamma$ is a discontinuous function of a parameter $\alpha$.

\vspace{0.5cm}

We consider the family $\{  P_{\infty}(\alpha) = (A$ ,$B(\alpha)$ ,
$p_1$ ,
$p_2)$, $ \alpha  \in  {\RR}^{+*}$, $p_1 > 0$ , $p_2 > 0 \}$ of
infinite Markovian  products of the singular matrix $A =
\pmatrix {1 & 0 \cr 0 & 0 \cr}$ and a family  $B(\alpha )$ of
matrices, depending on a parameter $\alpha$. $p_1$ and $p_2$ are the
transition probabilities defined by $(2.1)$.

\vspace{0.5cm}

If $p_1 + p_2 = 1$, then we recover the case of
a Bernoulli product.

$B(\alpha )$ is defined  by
$$B(\alpha ) = \Rt(-\varphi) \pmatrix{
\cosh {\sigma}_0                  & {\rho} \sinh {\sigma}_0 \cr {1
\over {\rho}}\sinh {\sigma}_0   &  \cosh {\sigma}_0  \cr}
\Rt(\varphi)\eqno(6.1)$$
where $\varphi  = -{\pi \over 4}$, $\sigma
> 0$ is a fixed parameter; and $\rho = e^{\alpha}$ with $\alpha \in
{\RR}^{+*}$.

\vspace{0.5cm}

The condition $b_{11}(\alpha ) = 0$ gives us
$$ \alpha = {\alpha}_n
= - Argch( \coth(n{\sigma})) \eqno(6.2)$$
and $\gamma ({\alpha}_n) =
-\infty$.

\vspace{0.5cm}

Now for all $\alpha \in {\RR}^{+*}$, we define
$$
\begin{array}{ll} f(\alpha)  & =
\gamma(\alpha) = {p_2 \over {p_1 + p_2}} {\sigma}_0 + {{p_1 p_2}
\over {p_1 + p_2}} \log({{\cosh\alpha  - 1} \over 2})  \\
 & +  {{p_1^2 p_2} \over {p_1 + p_2}} \sum_{n=1}^{\infty}
{(1-p_1)^{n-1} \log(1 + {{\cosh\alpha  + 1} \over {\cosh\alpha  -
1}} e^{-2n{\sigma}_0)}) }
\end{array} \eqno (6.3)$$
Since the series in $(5.3)$ is sommable, then  $f(\alpha )$ is defined
and continuous.

But since, for $ \alpha \neq {\alpha}_n$, we have
$$ \gamma ( \alpha ) = f( \alpha ),\eqno(6.4)$$

We obtain

$$\lim_{\alpha \rightarrow
{\alpha}_n}{\gamma(\alpha)} = f({\alpha}_n) \neq
\gamma({\alpha}_n)\eqno(6.5)$$
for all $n \in {\NN}^*$. Thus  $\gamma (\alpha )$ is a discontinuous
function of $\alpha$, for each  $\alpha = {\alpha}_n$, $n \in \NN$.

\end{document}